\documentclass[a4paper,fleqn,usenatbib]{mnras}

\usepackage{newtxtext,newtxmath}
\usepackage{url}

\usepackage[T1]{fontenc}
\usepackage{ae,aecompl}
\usepackage{pdflscape}
\usepackage{bm}
\usepackage{amsmath}

\usepackage{graphicx}	
\usepackage{amsmath}	
\usepackage{amssymb}	

\newcommand{\msun}{\,M$_{\odot}$}
\newcommand{\pyr}{\,yr$^{-1}$}
\newcommand{\au}{\,au}
\newcommand{\ps}{\,s$^{-1}$}

\title[Orbital evolution in symbiotic binaries]{Wind-accelerated orbital evolution in binary systems with giant stars}

\author[Zhuo Chen et al.]{
Zhuo Chen,$^{1}$\thanks{E-mail: zhuo.chen@rochester.edu}
Eric G. Blackman,$^{1}$\thanks{E-mail: blackman@pas.rochester.edu}
Jason Nordhaus,$^{2,3}$
Adam Frank$^{1}$\newauthor
and Jonathan Carroll-Nellenback$^{1}$
\\
$^{1}$Department of Physics and Astronomy, University of Rochester, NY 14627, USA\\
$^2$National Technical Institute for the Deaf, Rochester Institute of Technology, NY 14623, USA\\
$^3$Center for Computational Relativity and Gravitation, Rochester Institute of Technology, NY 14623, USA\\
}

\date{Accepted XXX. Received YYY; in original form ZZZ}

\pubyear{2017}

\begin{document}
\label{firstpage}
\pagerange{\pageref{firstpage}--\pageref{lastpage}}
\maketitle

\begin{abstract}
Using 3D radiation-hydrodynamic simulations and analytic theory, we study the orbital evolution of asymptotic-giant-branch (AGB) binary systems for various initial orbital separations and mass ratios, and thus different initial accretion modes. The time evolution of binary separations and orbital periods are calculated directly from the averaged mass loss rate, accretion rate and angular momentum loss rate. We separately consider  spin-orbit synchronized and zero spin AGB cases. We find that the the angular momentum carried away by the mass loss together with the mass transfer can effectively shrink the orbit when accretion occurs via wind-Roche-lobe overflow. In contrast,  the larger fraction of mass lost in Bondi-Hoyle-Lyttleton accreting systems  acts to enlarge the orbit. Synchronized binaries tend to experience stronger orbital period decay in close binaries. We also find that orbital period decay is faster when we account for the nonlinear evolution of the accretion mode as the binary starts to tighten. This can increase the fraction of binaries that result in common envelope, luminous red novae, Type Ia supernovae and planetary nebulae with tight central binaries. The results also imply that planets in the the habitable zone around white dwarfs are unlikely to be found. 
\end{abstract}




\section{Introduction}
Binary systems are progenitors for a wide range of astrophysical systems given the multiplicity of evolutionary outcomes. For much of a binary's lifetime, the component stars may be non-interacting.  However when one of the star evolves to the red giant branch (RGB) or asymptotic giant branch (AGB), the stars may interact via  wind-mass transfer and/or  tidal friction \citep{zahn1977}; their subsequent mutual evolution is strongly coupled \citep{paczynski1971}.

The mass transfer interaction can be classified into four types in increasing order of interaction. These are: Bondi-Hoyle-Lyttleton (BHL) accretion \citep{hoyle1939,bondi1944,edgar2004}; wind-Roche-lobe overflow (WRLOF) \citep{podsiadlowski2007}; Roche-lobe overflow (RLOF) \citep{paczynski1971}; and common envelope (CE) \citep{ivanova2013}.  Secular and dynamical instabilities can develop before the Roche limit  contact is reached, which may  in turn propel the system into the RLOF regime \citep{lai1994}.

The aforementioned mass transfer modes are commonly studied separately, but  they are often successive stages of time-evolving systems. For example, RLOF and tidal friction (TF) can shrink the orbit \citep{tout1991,lai1994} and may lead to CE when both stars' Roche lobes are filled. As we will see, at even larger initial separations, WRLOF can effectively transfer mass between binary stars while depositing angular momentum into a circumbinary disc. The system can then evolve to RLOF as the binary orbit shrinks. This exemplifies how  these three modes of mass transfer can be connected by orbital period decay (OPD). As we suggest herein, a sufficiently rapid evolution to the CE stage, even from  widely separated binaries,  implies that there may be many more binary systems that can ultimately arrive at CE than would be estimated using only their initial separation.

Example systems which result from  binary interactions for which our study is germane  include, luminous red novae (LRNe), type Ia supernovae (SNe), and planetary nebulae (PNe).  LRNe  have luminosities lower than typical supernovae that occur on white dwarfs (WD) but higher than novae. LRN light curves peak in the  optical before the infrared. Since 1989 \citep{rich1989}, many new LRNe have been observed: M31RV \citep{rich1989}, V4332 Sgr \citep{martini1999}, V838 Mon \citep{brown2002,bond2003,munari2005,tylenda2005}, M85 OT2006-1 \citep{kulkarni2007,rau2007}, V1309 Scorpii \citep{tylenda2011} and the recent M31LRN 2015 \citep{macleod2017}. Their origin is not fully understood;  some may be caused by a helium flash while others may result from merging binary systems \citep{pejcha2016,staff2016,metzger2017}. For the latter 'mergeburst' scenario, it is believed that the binary system will incur a CE \citep{nordhaus2006,ivanova2013}. During this phase, a considerable fraction of the envelope may be ejected or pushed to larger orbits. When the central binary sufficiently tightens or merges, kinetic energy will be released and the ejected envelope will be heated. Prior to the CE phase, there would also be a phase of RLOF. RLOF and CE phases also likely  precede Type Ia SNe \citep{iben1984,kenyon1993,ivanova2013,santander2015}. 

PNe are the nebular end states of low-mass stars \citep{han1995}. PNe have been observed to have a range of shapes, mostly aspherical, and often bipolar and asymmetric \citep{balick2002}.  Many PNe might be explained by binary models \citep{soker1994,mastrodemos1998,nordhaus2006,hillwig2016,chen2017,kim2017} but some might require triples \citep{bear2017}.Some PNe  have WD-WD binaries in the central region \citet{santander2015,miszalski2017}. In rapidly evolving binary system, magnetic fields can be an intermediary in the conversion of rotational energy into jets and asymmetric outflows \citep{nordhaus2007,nordhaus2011}. All of this highlights the importance of assessing how binary systems evolve as a function of initial conditions.

Another interesting context for our study is the effort to determine what kind of binary systems with planets or secondaries survive around WD.  \citet{villaver2009} and \citep{nordhaus2010} investigated the orbital change in 
low-mass binary systems and found that tidal friction, gravitational drag and mass loss from the primary are responsible for orbital change. Given the engulfment of low-mass companions during the giant phase of the primary, \citet{nordhaus2013} found that planetary companions will be tidally disrupted during the CE phase and thus not remain intact in close orbits around WDs. For these systems, engulfment by  CE is likely in the final stages of OPD, but may have been preceded by RLOF and WRLOF. RLOF and WRLOF are fundamentally different from BHL, the latter being the  mass transfer mode assumed by \cite{villaver2009}.

With CE being such a  key final stage determining the phenomenology of many stellar and planetary systems, a basic question is to understand how wide initial binary systems can be and  still arrive at CE. In particular what binary systems incur WRLOF and RLOF on the  path to CE? If the evolution through these stages is rapid enough (for example, shorter than an AGB stellar lifetime), then even systems with initial separations outside  of a CE, might evolve to CE. We explore this  in the present paper.  The results are important for improving  the statistics of binary evolution. 

Our study combines our previous 3D radiation hydrodynamic simulations of WRLOF and BHL accretion in AGB binary systems, with analytic theory. In Section \ref{sec:nmodel}, we briefly describe the numerical model of our simulations. In Section \ref{sec:amodel}, we present the analytic model of synchronized and non-synchronized binaries. 
In Section \ref{sec:main}, we apply the numerical results from simulations to the analytic model to characterize the orbital period change rate in realistic AGB binary systems. We then compare the results with both conservative mass transfer systems and ideal BHL mass transfer systems. In Section \ref{sec:imp}, we discuss the phenomenological implications. We conclude in Section \ref{sec:cons}.

\section{Numerical Model}\label{sec:nmodel}

A detailed description of the numerical model can be found in \citet{chen2017}. Here we outline the salient features for the present simulations.

The 3D radiation hydrodynamic simulations,  performed with \small{ASTROBEAR}\footnote{\url{https://astrobear.pas.rochester.edu/}} \citep{cunningham2009,carroll2013}, are carried out in the co-rotating frame of the binary systems (table \ref{tab:modellist}).  When the 
boundary of the giant star is stationary in the co-rotating frame, the giant star is in spin-orbit synchronized rotation in the lab frame (Appendix \ref{apdix:synced}). 
In addition, a 2D ray tracing algorithm, cooling and dynamic dust formation are considered in these simulations. The wind from the giant star is driven by a piston model at the inner boundary of the giant star and the radiation pressure where dust present. The secondary is accreting the gas \citep{krumholz2004}.

The numerical models replicate the BHL mode of mass transfer in binaries with large separation and WRLOF mode of mass transfer in close binary systems. The morphology of outflow is similar to some well known objects such as L2 Puppis \citep{kervella2016}, CIT 6 \citep{kim2017} and R Sculptoris \citep{maercker2012}.

Since the simulations are carried out in Cartesian coordinate with Eulerian code, an important concern is how well  angular momentum is conserved. We find 
that angluar momentum in the  wind to be conserved within $4\%$, up to 1.4 times  the orbital separation,  which is good enough not to affect  our conclusions. Detailed analysis can be found in Appendix \ref{apdix:error}.

Although the numerical simulations are carried out in the co-rotating frame as mentioned above, 
all of our analytic calculations are carried out for  the lab frame. We  designate subscript '1' to the giant star and subscript '2' to the secondary.

\section{Analytic Model}\label{sec:amodel}

\citet{boyarchuk2002} and  lecture notes by Pols \footnote{\url{http://www.astro.ru.nl/~onnop/education/binaries_utrecht_notes/Binaries_ch6-8.pdf}} provide general equations for calculating the evolution of binary separation. \citet{tout1991} and \citet{pribulla1998} also developed analytic models of orbital evolution of binary stars, considering mass transfer, mass loss and angular momentum loss. Here we follow a similar method to derive the orbital period change rate for both synchronized and non-spinning  binaries.
For present purposes "synchronized" refers to the spin-orbit synchronization of our  primary AGB  stellar rotation period with the orbital spin, as measured in the lab frame. We do not consider the spin of the secondary.

We assume a binary  with primary mass $m_{1}$,  secondary mass $m_{2}$, and orbital separation $a$. The z-component of angular momentum of in the lab frame  for a fully synchronized state can be expressed by:
\begin{equation}\label{eqn:jfull}
    J=\frac{m_{1}m_{2}(Ga)^{1/2}}{(m_{1}+m_{2})^{1/2}}+\frac{I_{1}(Gm_{1}+Gm_{2})^{1/2}}{a^{3/2}}+\frac{I_{2}(Gm_{1}+Gm_{2})^{1/2}}{a^{3/2}}
\end{equation}
where $I_{1}$ and $I_{2}$ are the moment of inertia of each star. $G$ is the gravitational constant.

For binary systems that consist of a giant star and a main sequence star or WD, the giant  will contribute most of the moment of inertia. The gas accreted onto the secondary  could spin it up and a rapidly spinning secondary could feed back  on the accretion. Such feed back  warrants  a detailed  theoretical/numerical model that includes magnetic field and radiation  \citep{springel2005}.
Although recognizing that this is important for further work, here  we ignore the spin and moment of inertia of the secondary. 
Equation \ref{eqn:jfull} then becomes
\begin{equation}\label{eqn:jsim}
    J=\frac{m_{1}m_{2}(Ga)^{1/2}}{(m_{1}+m_{2})^{1/2}}+\frac{\alpha_{1}m_{1}R_{1}^{2}(Gm_{1}+Gm_{2})^{1/2}}{a^{3/2}}
\end{equation}
Where $\alpha_{1}$ and $R_{1}$ are the moment of inertia factor and radius of the AGB star, respectively. The orbital period of the binary system can be expressed as
\begin{equation}\label{eqn:period}
    P=2\pi a^{3/2}(Gm_{1}+Gm_{2})^{-1/2}.
\end{equation}
Squaring Eqns. (\ref{eqn:jsim}) and (\ref{eqn:period}) and taking the time derivative gives
\begin{equation}\label{eqn:j}
    2J\dot{J}=A\dot{a}+B,
\end{equation}
and
\begin{equation}\label{eqn:p}
    2P\dot{P}=\frac{12(a\pi)^{2}\dot{a}}{G(m_{1}+m_{2})}-\frac{4\pi^{2}a^{3}(\dot{m}_{1}+\dot{m}_{2})}{G(m_{1}+m_{2})^{2}},
\end{equation}
where
\begin{equation}
\begin{split}
    A=&\left(\frac{Gm_{1}m_{2}}{\sqrt{Ga(m_{1}+m_{2})}}-\frac{3\alpha_{1}R^{2}m_{1}\sqrt{G(m_{1}+m_{2})}}{a^{5/2}}\right)\\
     &\times\left(\frac{\sqrt{Ga}m_{1}m_{2}}{\sqrt{m_{1}+m_{2}}}+\frac{\alpha_{1}R^{2}m_{1}\sqrt{G(m_{1}+m_{2})}}{a^{3/2}}\right)
\end{split}
\end{equation}
and
\begin{equation}
\begin{split}
    B=&2\times\left(\frac{\sqrt{Ga}m_{1}m_{2}}{\sqrt{m_{1}+m_{2}}}+\frac{\alpha_{1}R^{2}m_{1}\sqrt{G(m_{1}+m_{2})}}{a^{3/2}}\right) \\
      &\times\Bigg(\frac{\sqrt{Ga}m_{2}\dot{m}_{1}}{\sqrt{m_{1}+m_{2}}}+\frac{\alpha_{1}R^{2}\dot{m}_{1}\sqrt{G(m_{1}+m_{2})}}{a^{3/2}}+\frac{\sqrt{Ga}m_{1}\dot{m}_{2}}{\sqrt{m_{1}+m_{2}}}\\
      &-\frac{\sqrt{Ga}m_{1}m_{2}(\dot{m}_{1}+\dot{m}_{2})}{2\left(m_{1}+m_{2}\right)^{3/2}}+\frac{\alpha_{1}R^{2}G(\dot{m}_{1}+\dot{m}_{2})}{2a^{3/2}\sqrt{G(m_{1}+m_{2})}}\Bigg)
\end{split}
\end{equation}
Given $\alpha_{1},R_{1},\dot{J},\dot{m}_{1}$ and $\dot{m}_{2}$, $\dot{a}$ can be calculated with Eqn. (\ref{eqn:j}) and $\dot{P}$ can be calculated by feeding $\dot{a}$ into Eqn. (\ref{eqn:p}). If the primary has zero spin, taking $\alpha_{1}=0$  gives the appropriate result.

\section{Results Using Simulations to Fix Model Parameters }\label{sec:main}

\subsection{Extracting Parameters from Simulations }\label{sec:measure}
To measuring  $\dot{m}_{2}$, we simply keep track of the mass of the secondary in our simulations. For $\dot{m}_{1}$, we need to measure the mass loss from the binary system. We do so by summing up the flux through a spherical sampling shell centered at the  center of mass of the binary. The sampling shell is chosen large enough to contain both stars. The mass flux through the sampling shell is not sensitive to the size of the shell since \small{ASTROBEAR} uses a conservative scheme to conserve mass strictly. Therefore any
 convenient radius that is large enough will do
 \citet{chen2017}.

Choosing a sampling shell to calculate  $\dot{J}$ is more subtle. \citet{lin1977} found that the specific angular momentum of escaping parcels about the binary center of mass
will continue to increase when moving outward in the orbital plane. The parcel approaches its final specific angular momentum after it reaches $3$ times the binary separation $a$. \citet{macfadyen2008} found a  similar result. Therefore, spherical sampling shells with radii
$ r_\text{shell}\le  3a$ that contain both of the stars  may in general give different $\dot{J}$..

Another  concern is identifying from whence the escaping mass originates. Usually, mass loss from the L2 and L3 Lagrangian point is thought to be escaping mass.
But the Roche potential is not accurate for a very luminous binary system \citet{chen2017}. The L2 or L3 point may move inward or disappear when the luminosity increases. In our 3D simulation, the AGB star is pulsating periodically and the corresponding L2 point is oscillating. Thus there is no easy way to pinpoint 
specific radii which distinguish escaping mass and non-escaping mass, so the best that one can do is make it as large as possible.

Our use of AMR poses a competing constraint that limits how large we can choose the $\dot{J}$ sampling radius. On one hand, AMR  allows us to put more computational resources in the central region of the binary system where the mass transfer warrants extra resolution. On the other hand, we must use a coarse grid at large radii,
increasing the error there,  especially where  
inertial forces dominate at large radii. 
To minimize the angular momentum conservation error  (Appendix \ref{apdix:error}), we  use spherical sampling shells with $1.3$ times of binary separation, centered at the center of mass of the binary. Such sampling shells contain both L2 and L3 points predicted by standard theory in all of our binary models. Given that escaping gas will continue to gain angular momentum as it moves out,  this gives a lower bound on $\dot{J}$.

The mass flux through the sampling shell is given by
\begin{equation}
    \dot{m}_\text{esc}(t)=\oint\limits_{S(1.3a)}\rho\mathbf{v}\cdot\text{d}\mathbf{S},
\end{equation}
where $S$ is the surface of the sampling shell with $r_\text{shell}=1.3a$, $\rho$ and $\mathbf{v}$ are the local fluid density and velocity, respectively. and  $\dot{m}_{1}(t)=-\dot{m}_\text{esc}(t)-\dot{m}_{2}(t)$ by mass conservation. 

The total angular momentum flux is 
\begin{equation}
    \oint\limits_{S(1.3a)}(\rho\mathbf{r}\times\mathbf{v})\cdot\mathbf{\hat{z}}\mathbf{v}\cdot\text{d}\mathbf{S}=-\dot{J}(t),
\end{equation}
where $\mathbf{\hat{z}}$ is the unit vector in $z$ direction. This $\dot{J}$ for the gas includes both the spin angular momentum that came from  the AGB star and the angular momentum gain interaction with the secondary.
For one our  cases in  Section \ref{sec:discussion},
we wil deduct  the spin angular momentum from $\dot{J}$.
We  introduce a dimensionless number
\begin{equation}
    \gamma(t)=\frac{\dot{J}(t)}{j_{0}\dot{m}_\text{esc}(t)},
\end{equation}
where $j_{0}=m_{1}m_{2}(Ga)^{1/2}(m_{1}+m_{2})^{-3/2}$ is the specific orbital angular momentum of the binary and  $\gamma$ measures how efficient the escaping gas is in removing angular momentum.

Since $\dot{m}_{1},\dot{m}_{2},\dot{J}$ and $\gamma$ (see Appendix \ref{apdix:gamma}) are time dependent,  we  average them as follows before using $\dot{m}_{1}$ and $\dot{J}$ in Eqn. (\ref{eqn:j}):
\begin{equation}\label{eqn:avgm1}
    \dot{m}_{1}=\int_{t_{i}}^{t_{f}}\dot{m}_{1}(t)\text{d}t,
\end{equation}
\begin{equation}\label{eqn:avgm2}
    \dot{m}_{2}=\int_{t_{i}}^{t_{f}}\dot{m}_{2}(t)\text{d}t,
\end{equation}
\begin{equation}\label{eqn:avgj}
    \dot{J}=\int_{t_{i}}^{t_{f}}\dot{J}(t)\text{d}t,
\end{equation}
and
\begin{equation}\label{eqn:avggamma}
    \gamma=\frac{\int_{t_{i}}^{t_{f}}\dot{J}(t)\text{d}t}{j_{0}\int_{t_{i}}^{t_{f}}\dot{m}_\text{esc}(t)\text{d}t}=\frac{\dot{J}}{j_{0}(\dot{m}_{1}+\dot{m}_{2})},
\end{equation}
where $t_{i}$ and $t_{f}$ are the initial and final sampling time, respectively. In practice, we choose $t_{i}$ when the simulation becomes stable and $t_{f}-t_{i}>9$yr. It turns out that $\gamma$ is a more intuitive number than $\dot{J}$ for comparing the angular momentum loss efficiency between different models. Recovering $\dot{J}$ with Equation (\ref{eqn:avggamma}) is straightforward. Table \ref{tab:sample} lists $\dot{m}_{1},\dot{m}_{2}$ and $\gamma$.
\begin{table}
    \centering
    \begin{tabular}{c|c|c|c|c}
    \hline
    model & $\dot{m}_{1}$ & $\dot{m}_{2}$ & $\gamma$ & $r_\text{shell}$ \\
    & \msun\pyr & \msun\pyr & & \au\\\hline
    1 &  $-2.96\times10^{-7}$ & $1.16\times10^{-7}$ & $7.77\times10^{0}$ & 3.9 \\\hline
    2 & $-3.28\times10^{-7}$ & $1.20\times10^{-7}$ & $2.74\times10^{0}$ & 5.2 \\ \hline
    3 & $-2.21\times10^{-7}$ & $4.51\times10^{-8}$ & $1.01\times10^{0}$ & 7.8 \\ \hline
    4 & $-2.66\times10^{-7}$ & $8.28\times10^{-9}$ & $9.13\times10^{-1}$ & 10.4 \\ \hline
    5 & $-2.61\times10^{-7}$ & $5.85\times10^{-9}$ & $8.03\times10^{-1}$ & 13 \\ \hline
    \end{tabular}
    \caption{Measured parameters from 3D simulations. The first column lists the binary model number (Table \ref{tab:modellist}). $\dot{m}_{1}$ and $\dot{m}_{2}$ are the mass change rate of the primary and the secondary, respectively. $\gamma$ is the average number calculated by Equation (\ref{eqn:avggamma}). $r_\text{shell}$ is the radius of the sampling shell (centered at the center of mass of the two stars) through which the escaping flux is measured.}
    \label{tab:sample}
\end{table}

\subsection{Conservative versus BHL mass transfer models}
Conservative mass transfer is likely for very close binaries incurring  RLOF, while BHL mass transfer is more likely for  wide binaries. We simulated the intermediate regime of 3D WRLOF which can lead to fractions as high as ($\sim40\%$)  of the wind mass  transferred. As such, conservative and BHL mass transfer models represent the two bounding extremes for WRLOF  for comparison. The relevant model description for these two extremes be found in  standard textbooks so we only  discuss them briefly here. Wwe neglect the spin of the stars in this section.

\subsubsection{conservative mass transfer model}
Since we ignore the spin of the stars, $\alpha_{1}=0$ in Eqn. \ref{eqn:j}. And by the definition of conservative mass transfer,  $\dot{m}_{2}=-\dot{m}_{1}$. We use the  average value of $\dot{m}_{1}$ (Eqn. \ref{eqn:avgm1}) that we measured from our simulations. In conservative mass transfer, $\dot{J}_\text{con}=\gamma_\text{con}=0$ since there is no mass loss from the binary system. $\dot{P}_\text{con}$ will be used to denote the orbital period change in this model. The results are listed in Table \ref{tab:con}.
\begin{table}
    \centering
    \begin{tabular}{c|c|c|c}
    \hline
        model & $\dot{m}_{1}$ & $\dot{m}_{2}$ & $\dot{P}_\text{con}$ \\
          & \msun\pyr & \msun\pyr & yr\pyr \\ \hline
        1 & $-2.96\times10^{-7}$ & $2.96\times10^{-7}$ & $-3.69\times10^{-5}$\\ \hline
        2 & $-3.28\times10^{-7}$ & $3.28\times10^{-7}$ & $-6.43\times10^{-6}$\\ \hline
        3 & $-2.21\times10^{-7}$ & $2.21\times10^{-7}$ & $-7.95\times10^{-6}$\\ \hline
        4 & $-2.66\times10^{-7}$ & $2.66\times10^{-7}$ & $-1.48\times10^{-5}$\\ \hline
        5 & $-2.61\times10^{-7}$ & $2.61\times10^{-7}$ & $-2.02\times10^{-5}$\\ \hline
    \end{tabular}
    \caption{Conservative mass transfer model. The model number in the first column corresponds to each model in Table \ref{tab:modellist}. $\dot{m}_{1}$ is assumed to be the same as the result from 3D simulation. $\dot{P}_\text{con}$ is the orbital period change rate calculated by Equation (\ref{eqn:j}) and Equation (\ref{eqn:p}). $\gamma_\text{con}=0$ in this model.}
    \label{tab:con}
\end{table}

\subsubsection{BHL mass transfer model}
To  estimate  the orbital period change rate $\dot{P}_\text{BHL}$ in the BHL accretion scenario, we assume the stars do not affect each others structure (mass loss, density distribution etc.) and are not  spinning. In BHL accretion (for negligible sound speed) \citep{bondi1944,edgar2004}, the accretion rate is given by
\begin{equation}\label{eqn:bhacc}
    \dot{m}_\text{BHL}=\frac{4\pi G^{2} M^{2}\rho_{\infty}}{v_{\infty}^{3}},
\end{equation}
where $M,\rho_{\infty},v_{\infty}$ are the mass of the accreting star, density and wind speed respectively. Here $M$ is the mass of the accreting star and $M=m_{2}$ in our binary models. 

Crudely, we assume
\begin{equation}\label{eqn:bhlrho}
    \rho_{\infty}=\frac{\dot{m}_\text{iso}}{4\pi a^{2}v_\text{wind}},
\end{equation}
and
\begin{equation}\label{eqn:bhlv}
    v_{\infty}=\sqrt{v_\text{wind}^{2}+(2\pi a/P)^{2}},
\end{equation}
where we take $-\dot{m}_{1}=\dot{m}_\text{iso}=2.31\times10^{-7}$\msun\pyr\  and $v_\text{wind}=15$km\ps\ as calculated for our isolated AGB model (see \citet{chen2017} Sec. 2.2). Here $P$ is the period of the binary, calculated from Eqn. (\ref{eqn:period}). We then get $\dot{m}_{2}=\dot{m}_\text{BHL}$ (Table \ref{tab:bhl}) by Equation (\ref{eqn:bhacc},\ref{eqn:bhlrho}) and Equation (\ref{eqn:bhlv}). We assume that the escaping gas has an  angular momentum per unit mass equal to the specific orbital angular momentum of the primary. By definition, $\gamma_\text{BHL}$ can be expressed as
\begin{equation}
    \gamma_\text{BHL}=\frac{m_{2}}{m_{1}}.
\end{equation}

\begin{table}
    \centering
    \begin{tabular}{c|c|c|c}
        \hline
        model & $\dot{m}_{2}$ & $\gamma_\text{BHL}$ & $\dot{P}_\text{BHL}$ \\
          & \msun\pyr & & yr\pyr \\ \hline
        1 & $1.04\times10^{-9}$ & 0.1 & $1.93\times10^{-6}$ \\ \hline
        2 & $1.44\times10^{-8}$ & 0.5 & $1.61\times10^{-6}$ \\ \hline
        3 & $8.92\times10^{-9}$ & 0.5 & $3.23\times10^{-6}$ \\ \hline
        4 & $6.12\times10^{-9}$ & 0.5 & $5.20\times10^{-6}$ \\ \hline
        5 & $4.47\times10^{-9}$ & 0.5 & $7.45\times10^{-6}$ \\ \hline
    \end{tabular}
    \caption{BHL mass transfer model. $\dot{m}_{1}$ is assumed to be $-2.31\times10^{-7}$\msun\pyr in this model. $\dot{m}_{2}$ is the accretion rate. $\dot{P}_\text{BHL}$ is the orbital period change rate calculated by Equation (\ref{eqn:j}) and Equation (\ref{eqn:p}).}
    \label{tab:bhl}
\end{table}

\subsection{Synchronized versus zero spin scenarios}
 Secondaries close to their giant star primaries can spin up (down) the latter via tidal forces \citep{zahn1989} and  angular momentum can be transferred to the giant convective envelope. AGB stars have thick convective envelopes below their photospheres. 
 
 There is also subsonic
 turbulence \citep{freytag2017}  between the dust formation shell and photosphere. This turbulent region could transfer angular momentum  in close binaries. A gas parcel could  gain or lose angular momentum as it makes radial excursions from pulsations through a differentially rotating 
atmosphere.  Where dust forms, the gas will rapidly accelerate to  supersonic speeds. The angular momentum transfer by convection  will be diminishing while angular momentum transfer by gravity becomes dominant. As we have justified in Appendix \ref{apdix:synced}, the AGB star of  models 1-4 is likely to be spin-orbit  synchronized but not in model 5. Instead of carrying out detailed calculation of the time dependent synchronized AGB star, we add the zero spin case calculation for each binary model. This  provides the complementary extreme to the synchronized cases for WRLOF binary systems.

To prepare for our  calculation of  the zero spin cases, we need to quantify $\dot{J}$. Since the binary simulations we use to inform the analytic models were performed for only the synchronized state, we must deduct the spin angular momentum from the AGB star to study the zero spin cases. We do that by measuring the angular momentum flux of an isolated AGB star in the co-rotating frame (same as in Appendix \ref{apdix:error}). Due to numerical error, $\dot{J}_\text{all}$  varies slightly with $r$ in our simulation so we must pick a radius to use. We  take the measured angular momentum flux at $r_\text{shell}=0.8a$ as  $\dot{J}_\text{all}$ and $\gamma_\text{all}=\dot{J}_\text{all}/(j_{0}\dot{m}_\text{all})$. This is justified because in all five models, $r_\text{shell}=0.8a$ is beyond the dust formation shell and all the angular momentum transfer by convection happens beneath  that shell. Subscript 'all' denote that this is all the angular momentum flux of the AGB wind which includes  both the spinning boundary and any transfer from convection in the subsonic region.

On the other hand, we can calculate  $\gamma_\text{spin}$ of just the spinning boundary (a spherical shell) located below the photosphere analytically. This is a measure of specific of angular momentum on the shell of the giant star. The result is
\begin{equation}
    \gamma_\text{spin}=\frac{2\omega R_{1}^{2}}{3j_{0}}=\frac{2(m_{1}+m_{2})^{2}R_{1}^{2}}{3m_{1}m_{2}a^{2}},
\end{equation}
where $R_{1}=0.9$au, the same as our 3D simulation and \small{MESA} code result. Subscript 'spin' denote that this is only the angular momentum inherited from the synchronously spinning boundary of the AGB star. We list  $\gamma_\text{all}$ and $\gamma_\text{spin}$ in Table \ref{tab:gamma}.

\begin{table}
    \centering
    \begin{tabular}{c|c|c}
    \hline
    model & $\gamma_\text{all}$ & $\gamma_\text{spin}$\\ \hline
    1 & $3.13\times10^{0}$ & $7.26\times10^{-1}$ \\ \hline
    2 & $9.38\times10^{-1}$ & $1.52\times10^{-1}$ \\ \hline
    3 & $3.34\times10^{-1}$ & $6.75\times10^{-2}$ \\ \hline
    4 & $1.37\times10^{-1}$ & $3.80\times10^{-2}$ \\ \hline
    5 & $8.18\times10^{-2}$ & $2.43\times10^{-2}$ \\ \hline
    \end{tabular}
    \caption{All angular momentum in AGB wind ($\gamma_\text{all}$) and angular momentum with a spinning boundary ($\gamma_\text{spin}$).}
    \label{tab:gamma}
\end{table}

 Table \ref{tab:gamma} shows that  spin angular momentum is not entirely negligible in very close binary simulation \citep{akashi2015} but negligible in wide binaries. The table also shows that
gas  gains angular momentum after it is ejected from the inner boundary of the AGB star. In a real system, this could result from
a physical viscosity that transports angular momentum from  inner  to outer layers, especially in the subsonic region.
However, computationally, Eulerian codes  have a substantial numerical viscosity. 
that cannot be eliminated. 
Presently, we  ascribe the aforementioned angular momentum transport to this numerical viscosity rather than a physical effect even though there are physical effects in real systems (including tidal friction) which may supply large values.

To  isolate the effect of numerical viscosity and  small error in non-conservation of angular momentum, we compare four cases
in our subsequent use of the simulation data to inform the theory: (i) spin-synchronized, with just the boundary AGB spin removed from the simulation data
 (labelled by subscript "syn,exspin"), (ii) spin-synchronized, with all of the wind angular momentum 
the  wind angular momentum  before the dust formation region  removed  from the simulation data  (labelled by subscript "syn,exall")
(iii) zero initial spin of the AGB star
where the AGB star has no initial spin kept and just the boundary spin is removed from the data, but allowing the wind to 
accumulate angular momentum (labelled by subscript "non,exspin"), (iv) zero initial spin of the AGB star and any subsequent boundary or  wind  angular momentum before the dust formation radius is removed from the data (labelled by subscript "non,exall").
.


%

. 
We use $\dot{J}_\text{exall}=\dot{J}-\dot{J}_\text{all}$ and $\dot{J}_\text{exspin}=\dot{J}-\dot{J}_\text{spin}$ to calculate the orbital period change rate for cases (ii,iv) and (i,iii) respectively. We employ the analytic model of Section \ref{sec:amodel}, and set $\alpha=0.063$ to calculate $\dot{a}_\text{syn}$ and $\dot{P}_\text{syn}$ in the synchronized binary cases and let $\alpha=0$ to calculate $\dot{a}_\text{non}$ and $\dot{P}_\text{non}$ for the  zero spin binary cases. The results are shown  in Table \ref{tab:result}.

\begin{table*}
    \centering
    \begin{tabular}{c|c|c|c|c|c|c|c|c|c|c|}
    \hline
    mod & $P$ & $\dot{a}_\text{syn,exspin}$ & $\dot{a}_\text{syn,exall}$ & $\dot{a}_\text{non,exspin}$ & $\dot{a}_\text{non,exall}$ & $\dot{P}_\text{syn,exspin}$ & $\dot{P}_\text{syn,exall}$ & $\dot{P}_\text{non,exspin}$ & $\dot{P}_\text{non,exall}$ \\
    & yr & \au\pyr & \au\pyr & \au\pyr & \au\pyr & yr\pyr & yr\pyr & yr\pyr & yr\pyr \\\hline
    1 & 4.96 & {\color{red} $-1.26\times10^{-5}$} & $-7.82\times10^{-6}$ & $-1.04\times10^{-5}$ & $-6.51\times10^{-6}$ & {\color{red} $-3.06\times10^{-5}$} & $-1.87\times10^{-5}$ & $-2.51\times10^{-5}$ & $-1.55\times10^{-5}$ \\\hline
    2 & 6.53 & {\color{red} $-2.83\times10^{-6}$} & $-1.41\times10^{-6}$ & $-2.78\times10^{-6}$ & $-1.41\times10^{-6}$ & {\color{red} $-6.22\times10^{-6}$} & $-2.75\times10^{-6}$ & $-6.10\times10^{-6}$ & $-2.73\times10^{-6}$ \\ \hline
    3 & 12.01 & {\color{red} $1.15\times10^{-7}$} & $5.91\times10^{-7}$ & $9.84\times10^{-8}$ & $5.69\times10^{-7}$ & {\color{red} $1.23\times10^{-6}$} & $2.66\times10^{-6}$ & $1.18\times10^{-6}$ & $2.59\times10^{-6}$ \\ \hline
    4 & 18.48 & $3.70\times10^{-7}$ & $6.53\times10^{-7}$ & {\color{red} $3.54\times10^{-7}$} & $6.34\times10^{-7}$ & $2.92\times10^{-6}$ & $3.90\times10^{-6}$ & {\color{red} $2.87\times10^{-6}$} & $3.84\times10^{-6}$ \\ \hline
    5 & 25.83 & $7.81\times10^{-7}$ & $9.81\times10^{-7}$ & $7.66\times10^{-7}$ & {\color{red} $9.66\times10^{-7}$} & $5.27\times10^{-6}$ & $6.05\times10^{-6}$ & $5.21\times10^{-6}$ & {\color{red} $5.98\times10^{-6}$} \\ \hline
    \end{tabular}
    \caption{The first column lists the binary model number; $P$ lists the orbital period in yr of each binary model. The binary separation change rate $\dot{a}$ and orbital period change rate $\dot{P}$ of four scenarios described in section 4.3 are listed in this table. \{syn,exspin\} refers to the model with a synchronized AGB star and for which we deduct the angular momentum inherited from the spinning inner boundary.  \{non,exall\}
     refers to the model that has non-spinning AGB star and we deduct all the angular momentum of the wind before it reaches the dust formation region. \{syn,exall\} and \{non,exspin\}  have the complementary analogous meanings. The red colors indicate the output from the models that we think best  characterize the physical expectation for the initial binary parameters. See text}
    \label{tab:result}
\end{table*}

Of the  four sets of angular momentum prescription choices in Table \ref{tab:result} that compensate to various degrees for the numerical error, 
we delineate the ones that we think best fit the physical binary model in red color. 
We base our judgment on the time scales listed in Table \ref{tab:modellist}. Specifically, we expect the AGB star in models 1-3 to be  synchronized  and the companion's tidal force can transfer angular momentum in the subsonic wind region. Model 4 has a longer synchronization timescale so  we expect that the inner part of the AGB star is not fully synchronized but that angular momentum transfer can
still be transferred within the  subsonic region. Model 5  probably would likely not correspond to a  synchronized state  and so we deduct all angular momentum in the AGB wind in that case.

\subsection{Results and physical discussion of orbital evolution}
\label{sec:discussion}

\begin{figure}
    \centering
    \includegraphics[width=1.0\columnwidth]{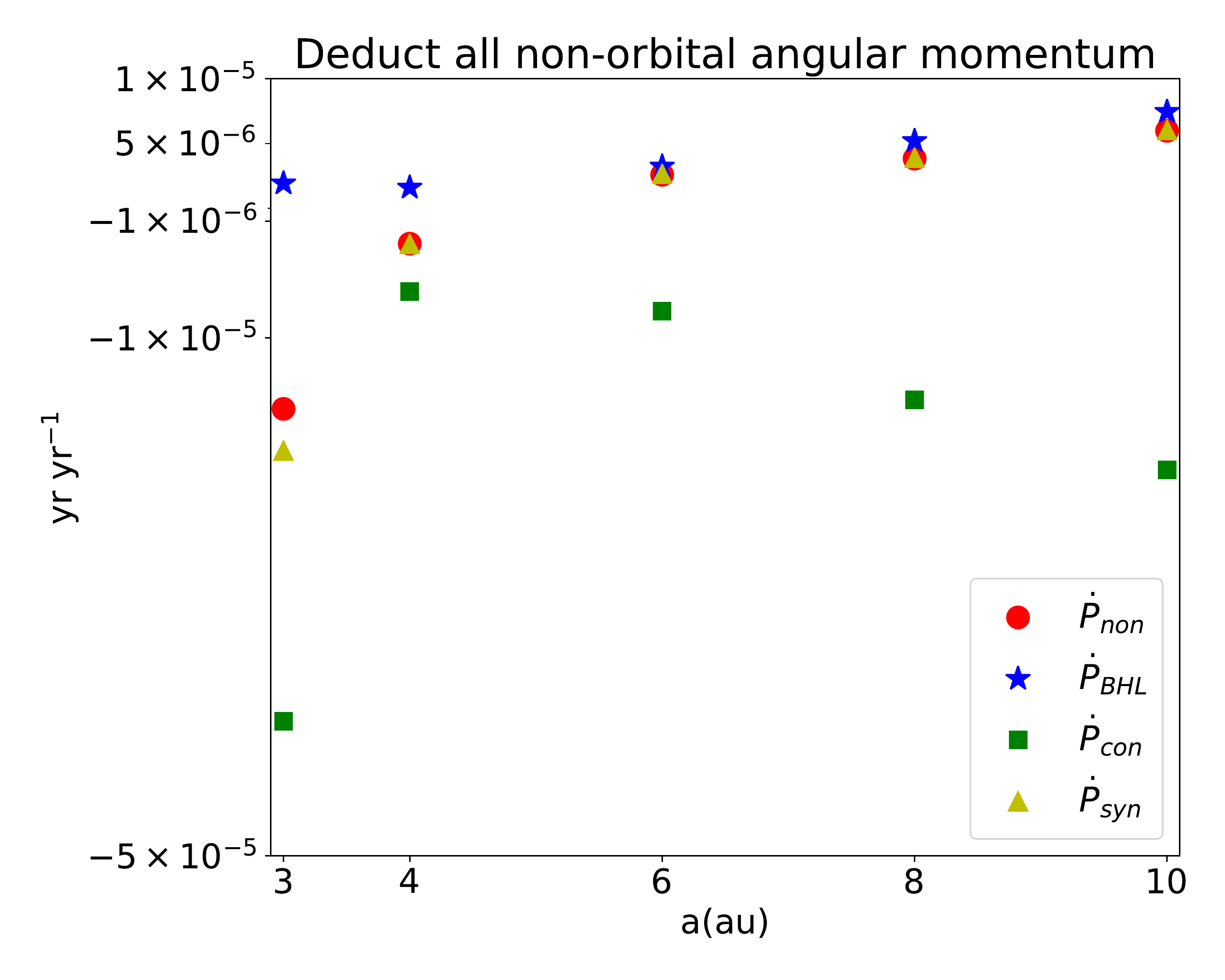}
    \includegraphics[width=1.0\columnwidth]{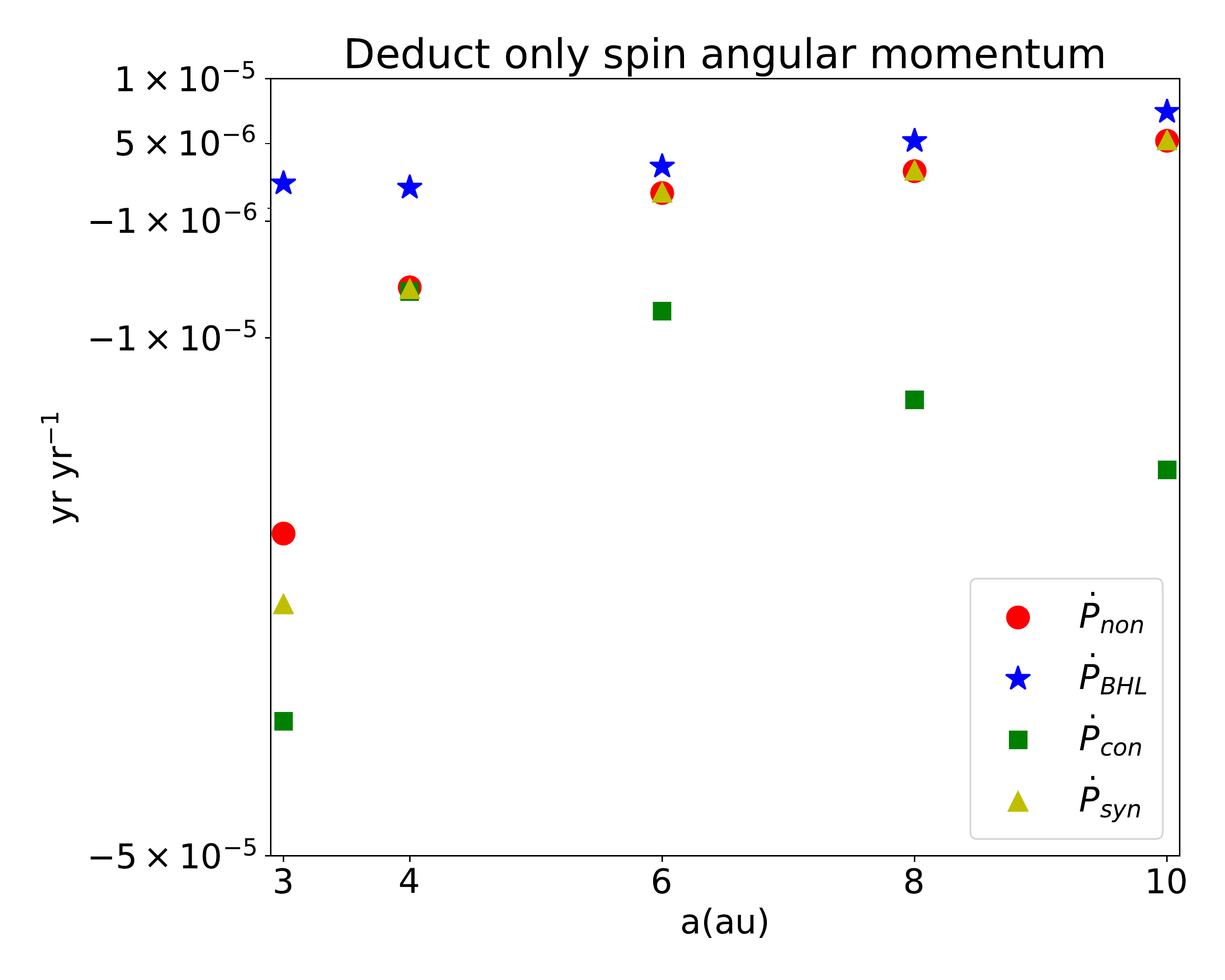}
    \caption{The first figure shows  $\dot{P}_\text{syn,exall}$ and $\dot{P}_\text{non,exall}$ for our five binary models vs. their initial orbital separation. The second figure shows the corresponding plot for $\dot{P}_\text{syn,exspin}$ and $\dot{P}_\text{non,exspin}$. $\dot{P}_\text{BHL}$ and $\dot{P}_\text{con}$ are calculated from our analytic model. Their exact values can be found in Table \ref{tab:result}.}
    \label{fig:dpdt}
\end{figure}

Results from \small{ASTROBEAR} and analytic models are combined in 
Figure \ref{fig:dpdt}. The plots show the  orbital period derivative for all of the different models as computed from their initial orbital
parameters.

As identified from Table \ref{tab:result} and Figure \ref{fig:dpdt}, two factors are most important: binary separation $a$ and the presence or absence of synchronization. We discuss each in turn.

In our 3D simulations, the AGB 
wind is standardized and so
the  binary separation $a$ will determine the mass transfer mode for a fixed secondary  mass. The binary stars of models 1-3 are likely experiencing WRLOF while those of models 4 and 5 are likely experiencing BHL mass transfer. The orbital period change rates $\dot{P}_\text{non}$ and $\dot{P}_\text{syn}$  approach $\dot{P}_\text{con}$ for small  binary separation 
but approaches $\dot{P}_\text{BHL}$ for large  separation (Figure \ref{fig:dpdt}). The trend is monotonic since $\dot{P}_\text{non}$ and $\dot{P}_\text{syn}$  both increase with  $a$. 
The orbital period decay (Table \ref{tab:result}) occurs rapidly fast enough in models 1 and model 2 that they will incur RLOF or precursor (Darwin) instabilities \citep{lai1994} within the lifetime of the AGB star ($\sim10^{6}$yr).
They will ultimately incur a CE phase when the companion dives into the AGB envelope.

The AGB wind provides the driving force that changes the orbital period is the AGB wind.  In its absense, there is no way to couple  spin and orbital angular momenta.
Without a wind  $\dot{J}=\dot{m}_{1}=\dot{m}_{2}=0$ giving $\dot{a}=0$ and $\dot{P}=0$ and there will be no difference in orbital evolution for synchronized and initially non-spinning models.
Comparing $\dot{P}_\text{syn,exspin}$ to $\dot{P}_\text{non,exspin}$ and $\dot{P}_\text{syn,exall}$ to $\dot{P}_\text{non,exall}$, we see that if $\dot{P}$ is negative, the synchronization makes it more negative. If $\dot{P}$ is positive, synchronization makes it more positive. This can be  understood if we view the rotating giant star as a reservoir of angular momentum that extracts from
the orbit  when the binary stars orbit faster and releases to the orbit  when the orbit is slower. Via the wind,  the spin-orbit synchronization of the giant star can thus accelerate  tightening or looseing  of  the orbit, the effect being stronger for closer binaries.

The wind speed also has some  influence on the rate of orbital shrinking.  
Faster winds have less time  to interact with the companion, thereby  reducing  angular momentum and mass transfer and increasing mass loss. 
From  our isolated AGB star model \citep{chen2017}, the terminal wind speed is $15$km\ps\ and the mass-loss rate is $2.31\times10^{-7}$\msun\pyr. Some AGB stars may eject slower winds ($5-10$km\ps) with higher mass-loss rates ($\sim10^{-6}$\msun\pyr) \citep{freytag2017}. In those cases, the binary system will be more likely to incur WRLOF 
the orbit will decay faster.

\section{Astrophysical implications}\label{sec:imp}

A most interesting consequence of our analysis is the implications for orbital period decay, which in turn leads strongly interacting binaries \citep{nordhaus2006,ivanova2016,staff2016} such as CE and tidal disruption. From Table \ref{tab:result} (and also Figure \ref{fig:dpdt}), we conclude that $\dot{P}$ is a highly non-linear function of $a$ but smaller $a$ leads to faster OPD. The non-linearity is   ascribed to the mode of mass transfer. Models 4 and 5 have BHL accretion which leads to 
relatively less interaction between the AGB wind and the secondary compared to  WRLOF or RLOF. In contrast, models 1, 2 and 3 have WRLOF which is a more effective mass transfer mechanism than BHL accretion. As such, the gas in the  can gain more angular momentum from the secondary and escape from the L2 point. When leaving the system, the gas carries a large fraction of angular momentum with a small fraction of mass of the binary as shown with $\gamma$ in Table \ref{tab:sample}.

As $\dot{a}$ increases monotonically with $a$ from negative  to positive  (Table \ref{tab:result}),
we can identify two key values $a_\text{merge}$ and $a_\text{bi}$. The former, $a_\text{merge}$, refers to the initial 
separation that distinguishes the binary systems that will and will not subsequently merge. The value $a_\text{bi}$ is the separation that distinguishes tightening from widening binaries. Specifically, in our simulation, $4$\au$<a_\text{merge}<a_\text{bi}<6$\au. The existence of such a boundary was previously inferred from a tidally interacting and BHL accretion binary model \citep{villaver2009,nordhaus2010} and likely exists for any low mass ratio two-body system.

\begin{figure}
    \centering
    \includegraphics[width=1.0\columnwidth]{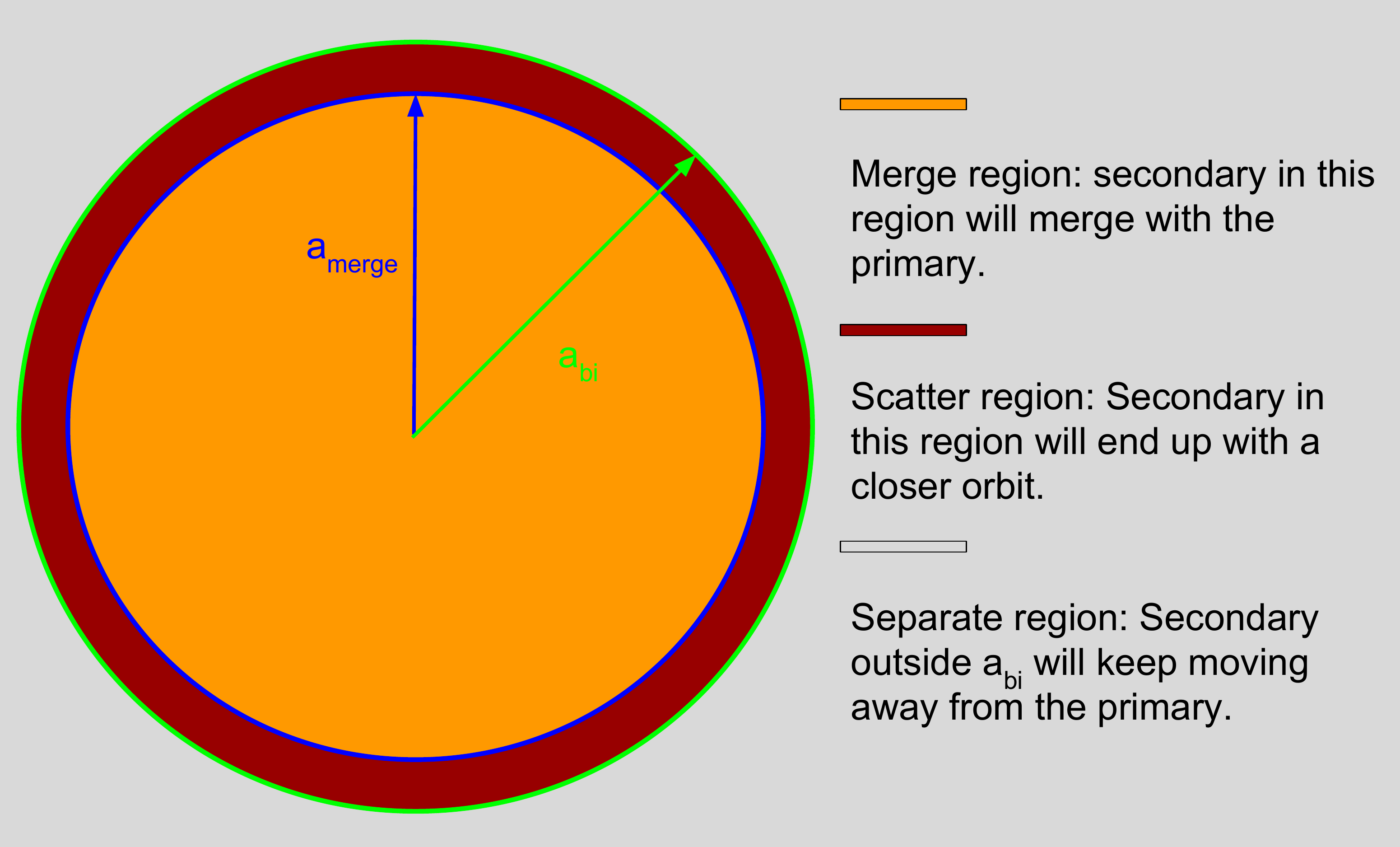}
    \caption{Conceptual fate of the secondary considering wind and tide.}
    \label{fig:fate}
\end{figure}

The fate of the secondary is summarized in Fig. \ref{fig:fate}. The primary is located at the concentric center of these spheres. The orange color indicates the merge region. A secondary in this region will move towards the primary due to OPD and incur CE. Red indicates the scatter region. This region is identified by $a_\text{merge}$ and $a_\text{bi}$. The secondary in this region will experience OPD but the OPD is not strong enough for merging to happen during the AGB lifetime. At the end of the AGB evolution, the position of the secondary will then be scattered within $a_\text{bi}$. 

\begin{figure}
    \centering
    \includegraphics[width=1.0\columnwidth]{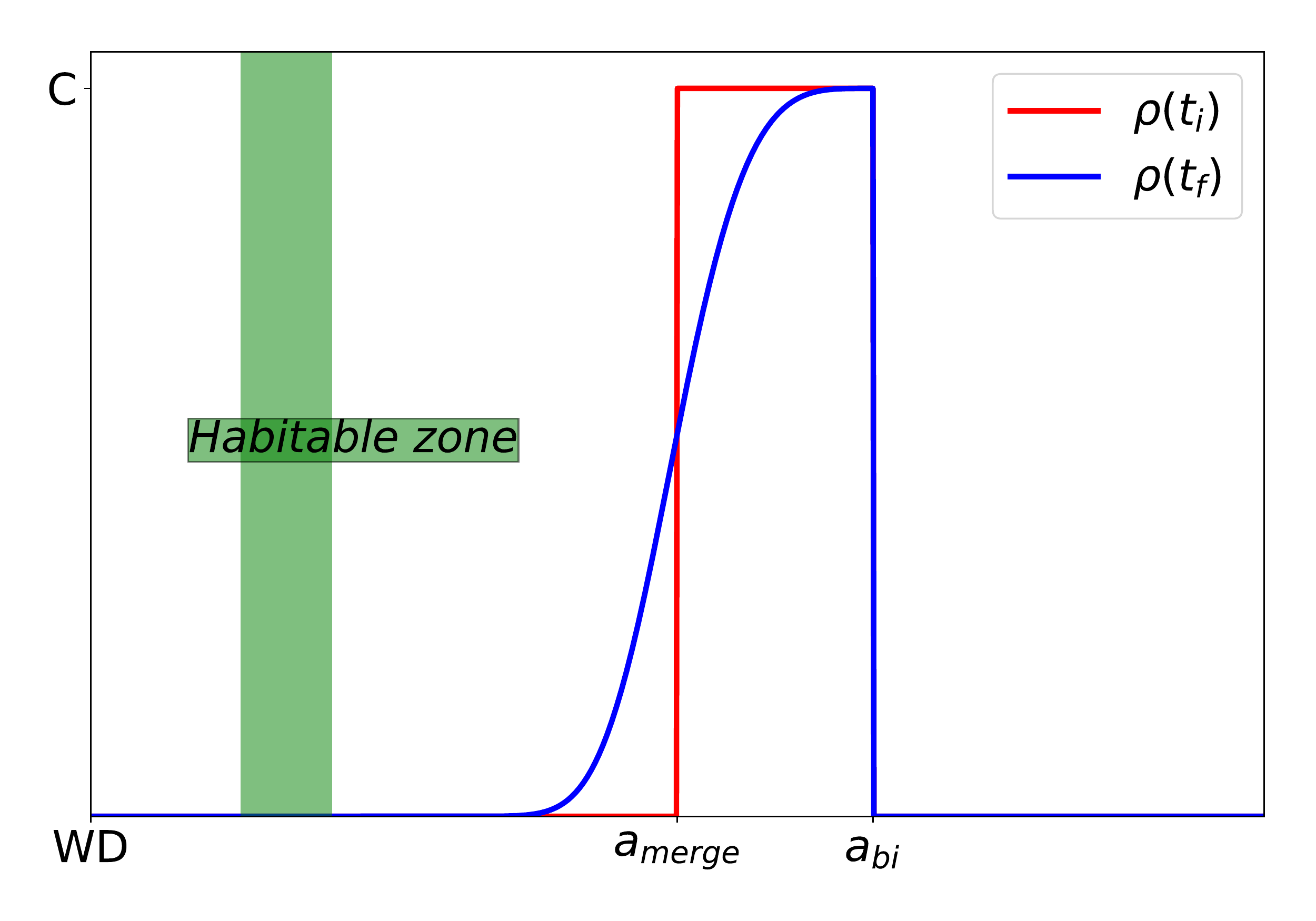}
    \caption{Conceptual $\rho(a,t_\text{i})$ and $\rho(a,t_{f})$. Green band represents the potential habitable zone and the WD is at the origin.}
    \label{fig:prob}
\end{figure}

Now we consider an ensemble of binary systems with secondaries  distributed uniformly only in the scatter region i.e.
\begin{equation}
    \rho(a,t_\text{i})=C
\end{equation}
where $\rho$ is  of being  initially
at binary separation $D(t_\text{i})=a$, where  $a_\text{merge}<a<a_\text{bi}$  at initial time $t_\text{i}$. Here $C$ is a constant to normalize the probability integral. Since $\dot{P}$ decreases rapidly with evolving $D$, when the binary system evolves to its final state at $t_\text{f}$, the closer the secondary is to the WD, the smaller the probability $\rho(D,t_\text{f})$. We illustrate $\rho(a,t_\text{i})$ and $\rho(a,t_\text{f})$ conceptually in Fig. \ref{fig:prob}.

Although not simulated directly, our analytic model, along with the three regions of Figure \ref{fig:fate}, also speaks to the question of  finding  potential planets around WDs \citep{nordhaus2013} and particularly, in the habitable zone. Surviving planets must be in the scatter region about the primary 
in order to migrate close enough but avoid CE, wherein it would be destroyed \citep{nordhaus2013}. However, this safety region is just thin shell of radial extent $<2$\au. Furthermore, $a_\text{merge}$ is usually not so small  in giant binary systems and the nonlinear evolution of $\dot{a}$ 
even if initially beyond the merge region, would  further decrease the possibility of finding a close planet around WD. This strengthens the argument that first-generation planets in the white dwarf habitable zone are unlikely unless tertiary or multi-body interactions result in scattering to high-eccentricity orbits \citep{nordhaus2013}.  Subsequent damping to circular orbits via tidal friction may be possible, but would initially leave rocky planets as uninhabitable, charred embers \citep{nordhaus2013}.

\section{Conclusions}\label{sec:cons}

We have studied the orbital period evolution in binary systems with a wind-emitting giant star, taking into account different modes (WRLOF and BHL accretion) of mass transfer ($\dot{m}_2$), mass loss ($\dot{m}_{1}$) and angular-momentum loss ($\dot{J}$). The time derivative of orbital separation $\dot{a}$ and that of the period $\dot{P}$ can be calculated analytically given $\dot{m}_{1},\dot{m}_{2}$ and $\dot{J}$. We have examined the validity if the giant star in our binary models can be synchronized under tidal force in Appendix \ref{apdix:synced} and considered both spin-orbit synchronized and zero spin AGB star scenarios.


We find that  giant stars in close binary systems that undergo WRLOF are likely to be  synchronized. We expect such systems to incur OPD and orbit decay to the point that RLOF or Darwin instabilities drive the system to a CE phase. Because WRLOF can happen at larger radii than RLOF or Darwin instabilities,the rapid evolution we find from the WRLOF phase implies an dramatic  increase in  the fraction of binaries that will arrive at CE were this phase ignored. In contrast, we find that wide binaries  undergo BHL mass transfer and are likely to be separating. In this case a synchronized giant star could serve as an angular momentum reservoir that further enhances the separation.

We have identified two characteristic binary separations $a_\text{merge}$ and $a_\text{bi}$: $a_\text{merge}$ is the initial critical separation below with binaries merge and $a_\text{bi}$ is the critical separation at which $\dot{a}=0$. These two separations divide the space into three regions (Fig. \ref{fig:fate}): merge, scatter, and separate.

Angular-momentum loss ($\dot{J}$) and mass loss ($\dot{m}_\text{esc}$) are two competing factors in our model, just as tidal friction, drag force and mass loss are in \citet{villaver2009,nordhaus2013}. All of these models agree that smaller separations lead to faster OPD although the mechanisms studied are different.

Finally, we emphasize the importance of 3D binary simulations for capturing the nonlinear evolution of the binary separation. In general, binary separation, mass of the stars and wind properties determine the mass transfer mode (RLOF, WRLOF, or BHL) and interaction (tides and instabilities). As binary stars get closer, the more a self-consistent giant star model is needed to resolve the fluid motion and stellar structure. 
The dynamical evolution between
mass transfer modes 
is  warrants 3D binary simulations. 
Crudely assuming only BHL accretion, for example, without following the nonlinear evolution of the accretion mode can miss the rapid OPD and subsequent merger if the actual system evolves to WRLOF when the separation decreases.

\section*{Acknowledgements}

We acknowledge support from grants HST-AR-13916.002 and NSF-AST1515648. We sincerely thank the anonymous referee who gave many valuable suggestions. ZC is grateful to Prof. Dong Lai and Prof. David Chernoff for providing place to write this paper. EB also acknowledges the Kavli Institute for Theoretical Physics (KITP) USCB
and associated support from grant NSF PHY-1125915. JN acknowledges support from NASA grants HST AR-14563 and HST AR-12146, and the National Technical Institute for the Deaf under Grant No. SPDI-15933.



\bibliographystyle{mnras}
\bibliography{mnras} 

\appendix
\section{Justifying Spin-Orbit Synchronized AGB star}\label{apdix:synced}
In our 3D simulations, the boundary condition of the AGB star we used is in spin-orbit synchronized state. However, the AGB stellar spin may not be synchronized in widely separated binaries with small mass ratios. To justify the assumption of synchronization, we calculate the synchronization timescale for the giant star using equation (5) in \citet{nordhaus2013}. If the timescale is much shorter than the lifetime of the AGB star and the timescale for the binary to merge, then the synchronized boundary condition can be justified.  The time-evolution of the spin is given by
\begin{equation}\label{eqn:domegadt}
    \left(\frac{\text{d}\Omega_{1}}{\text{d}t}\right)_{\text{tides}}=\frac{6\omega k_{2,1}f}{\alpha_{1}\tau_{1,\text{conv}}}\left(\frac{m_{1,\text{env}}}{m_{1}}\right)\left(\frac{m_{2}}{m_{1}}\right)^{2}\left(\frac{R_{1}}{a}\right)^{6}\left(1-\frac{\Omega_{1}}{\omega}\right)-\Omega_{1}\frac{\dot{I}_{1}}{I_{1}}
\end{equation}
where $\omega,\Omega_{1},R_{1},m_{1},m_{1,\text{env}},I_{1}$ are the circular orbital angular frequency of the binary system, rotational angular frequency, radius, total mass, envelop mass and moment of inertia of the giant star, $m_{2}$ is the total mass of the secondary, $\tau_{1,\text{conv}}=\left(m_{1,\text{env}}R_{1}^{2}/L_{1}\right)^{1/3}$ and $L_{1}$ is the luminosity of the giant star. The tidal Love number of the giant is $k_{2,1}$ which we assume to be unity, $f$ is also close to unity \citep{nordhaus2013}. For a spherically symmetric density distribution, the moment of inertia is calculated by:
\begin{equation}
    I=\int_{0}^{R}\frac{8\pi r^{4}\rho(r)}{3}\text{d}r=\alpha m R^2
\end{equation}
where $\alpha,R$ and $m$ are the moment of inertia factor, radius of the sphere (star) and the total mass of the star, respectively. In an evolving star, $I,\alpha,R$ and $m$ will all be time dependent variables. However, for simplicity, we only consider the time dependence of $I$ and $m$ in this paper and assume $\alpha$ and $R$ to be constant in a short period of time. 
The time scale for the giant star to be synchronized is then estimated by:
\begin{equation}
    t=\frac{\omega}{\text{d}\Omega/\text{d}t}
\end{equation}
taking $\Omega(t=0)=0$.

The remaining parameters needed to calculate $\text{d}\Omega/\text{d}t$ are: $\alpha,m_{1,\text{env}},R_{1}$ and $L_{1}$. \small{MESA} (Modules for Experiments in Stellar Astrophysics) \citep{paxton2015} can provide us the values. We take a 1.3\msun\ ZAMS as an illustrative example. When it evolves to 1\msun, $\alpha\approx0.063$, $m_{\text{env}}\approx0.45\text{M}_{\sun}$, $R_{1}\approx0.9\text{au}$ and $L\approx3100\text{L}_{\sun}$. We notice that the luminosity predicted by \small{MESA} is greater than the luminosity ($2342\text{L}_{\sun}$) we used in our 3D simulation. However, our AGB star model is a phenomenological model and we focus on the wind structure instead of the core and the envelope of the star. Our model has produced an AGB wind with reasonable speed ($\sim15\text{km/s}$) and a lower density. We view the discrepancy as a potential challenge and opportunity for future studies. To estimate the timescale (table \ref{tab:modellist}) of synchronization, we use the values from \small{MESA}.

From the resulting calculation,  we find that the timescale to synchronize the giant star is much smaller than the lifetime of the AGB star ($\sim10^{6}$yr) and the timescale to merge ($10^{5}-10^{6}$yr) (perhaps except for model 5) as we  see in the  calculations of section \ref{sec:discussion}). Therefore, we find it reasonable to assume that the AGB star should be synchronized throughout the simulations for models 1-3. On the other hand, we discuss and the non-synchronized nature of model 5 in more detail and give the corrected answer. The giant star in model 4 is likely partially synchronized.

\begin{table}
    \centering
    \begin{tabular}{||c|c|c|c|c||}
    \hline
    model & $m_{1}$ & $m_{2}$ & $a$ & $t$ \\
      & M$_{\sun}$ & M$_{\sun}$ & au & yr \\\hline
    1 & 1.0 & 0.1 & 3 & 2.41$\times10^{3}$ \\ \hline
    2 & 1.0 & 0.5 & 4 & 5.42$\times10^{2}$ \\ \hline
    3 & 1.0 & 0.5 & 6 & 6.17$\times10^{3}$ \\ \hline
    4 & 1.0 & 0.5 & 8 & 3.47$\times10^{4}$ \\ \hline
    5 & 1.0 & 0.5 & 10 & 1.32$\times10^{5}$ \\ \hline
    \end{tabular}
    \caption{The first column lists the model number used throughout this paper; $m_{1}$ and $m_{2}$ are the mass of the giant star and the secondary, respectively; $a(\text{au})$ is the binary separation; $t$ is the synchronization timescale of the giant star.}
    \label{tab:modellist}
\end{table}

As the binary separation is decreasing in model 1 and model 2, the synchronization assumption can hold for future long-term evolution (Perhaps model 3 also as the two stars are not separating fast). We will discuss the non-synchronous scenario in Section \ref{sec:discussion}.

\section{angular momentum error in AGB wind}\label{apdix:error}
In this section, we quantify the angular momentum error ('the error' hereafter) in our isolated AGB wind model. Ideally, when supersonic fluid is moving outward in a central potential (a combination of gravitational force and radiation force), its angular momentum should not change ($\dot{J}=$const.). In our model, gas will become supersonic when dust forms.

Since we have five binary models, each with different binary orbital angular frequencies, we examine the error of an isolated AGB wind for different angular frequencies. Each simulation has an AGB star at the center of the simulation box. The simulation is also carried out in co-rotating frame, with $\omega=\sqrt{G(m_{1}+m_{2})/a^{3}}$. This should be equivalent to simulation of AGB star which spins at $\omega$ in lab-frame. We measure the average angular momentum flux $\dot{J}$ through sampling shells (centered at the center of the AGB star) at a series of radii $[0.8a,0.9a,1.0a,1.1a,1.2a,1.3a,1.4a]$. $a$ is the binary separation.

\begin{figure}
    \centering
    \includegraphics[width=1.0\columnwidth]{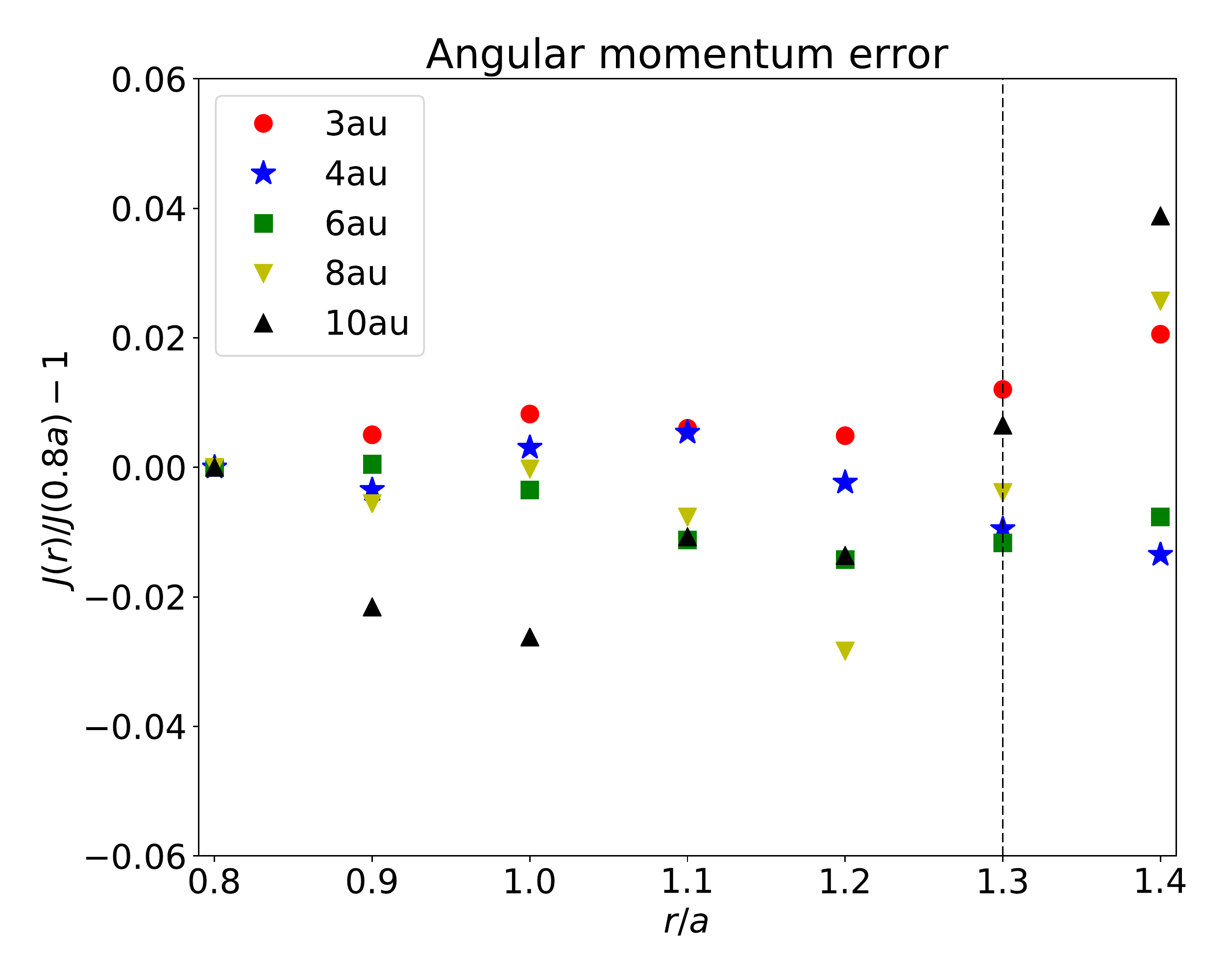}
    \caption{Angular momentum error in AGB wind. The x-axis is the distance in binary separation. The y-axis is the value of error. Five different makers show the error of corresponding angular frequency of the binary model with that separation. The vertical dotted line denote the radius of sampling shell that we use in parameter measurement (Section \ref{sec:measure}).}
    \label{fig:amerror}
\end{figure}
We plot the error of different models in Figure \ref{fig:amerror}. We can see that the error in AGB wind is acceptable, or within $4\%$ of all the angular momentum in the AGB wind.

\section{time varying $\gamma$}\label{apdix:gamma}

The AGB binary models in this paper take dust formation, radiation transfer and cooling into consideration and thus the system is highly dynamic, especially when there is a large accretion disc. We note that Eulerian codes in Cartesian coordinates are generally poor at conserving angular momentum at large  amid the coarse grid. We thus  we measure the angular momentum flux at relatively small radii $r_\text{flux}=1.3a$. The AMR capability of {\small ASTROBEAR} can help us resolve the central part of the binary system and thus minimize the error.

\begin{figure}
    \centering
    \includegraphics[width=1.0\columnwidth]{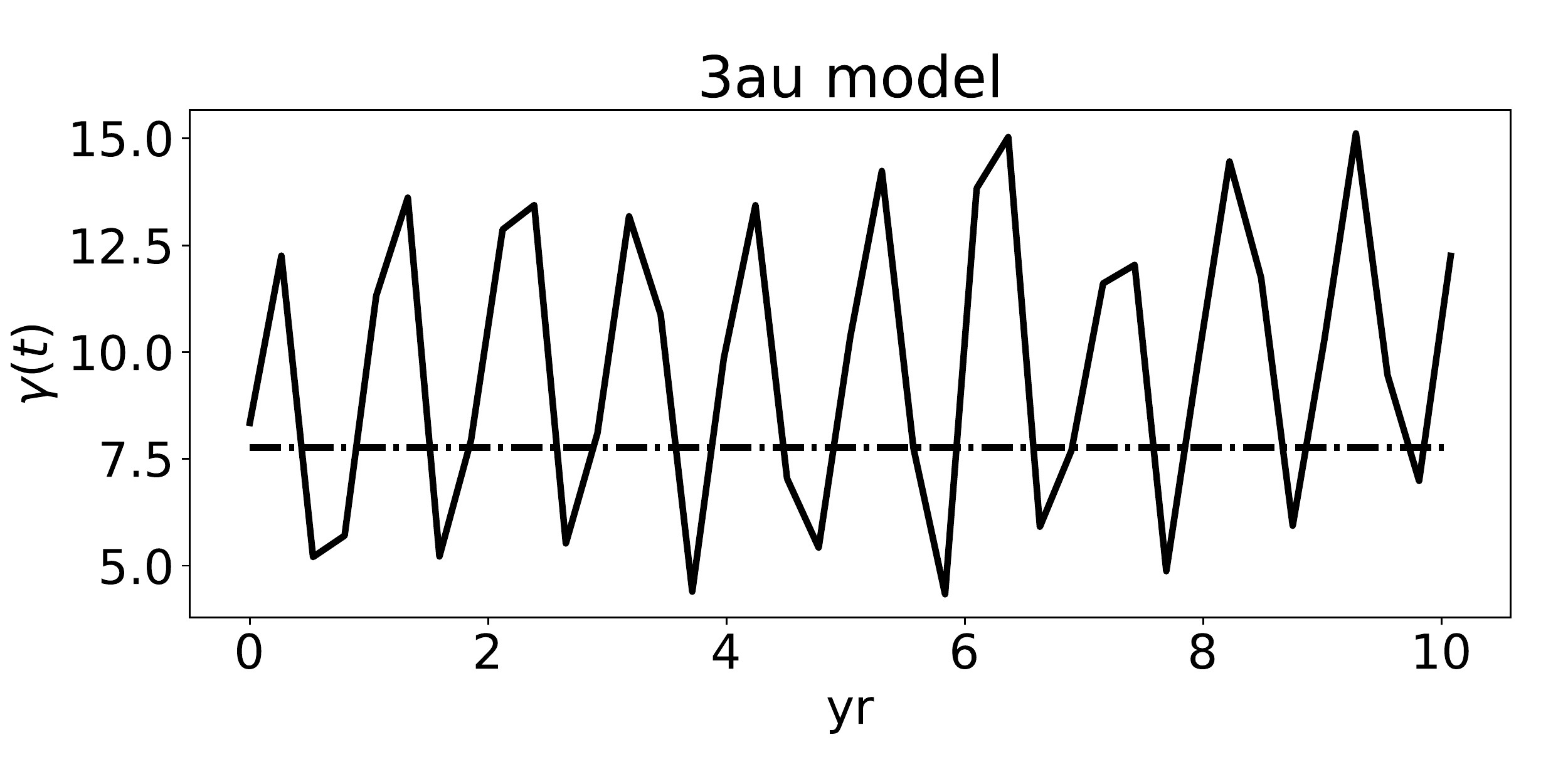}\\
    \includegraphics[width=1.0\columnwidth]{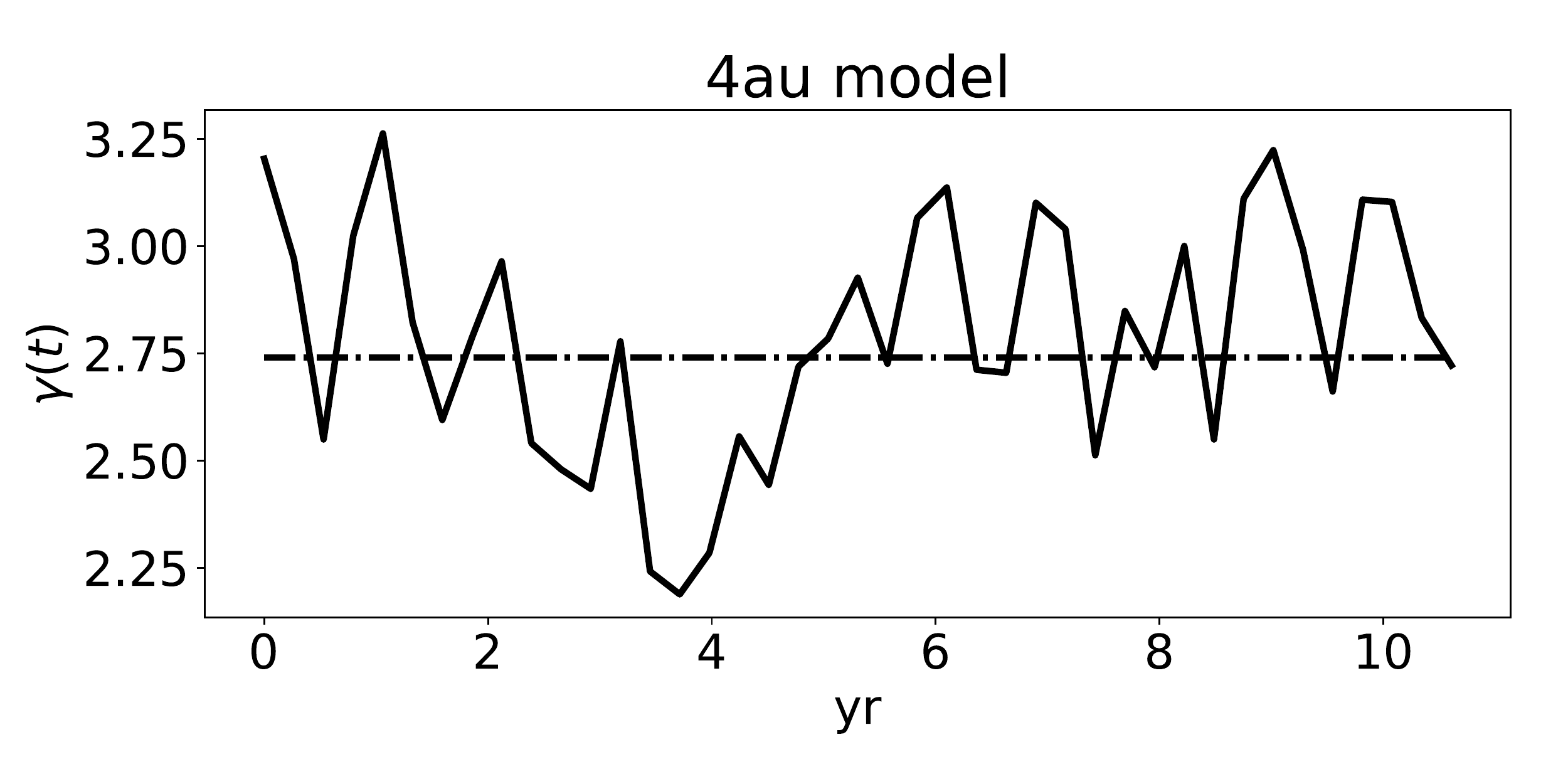}\\
    \includegraphics[width=1.0\columnwidth]{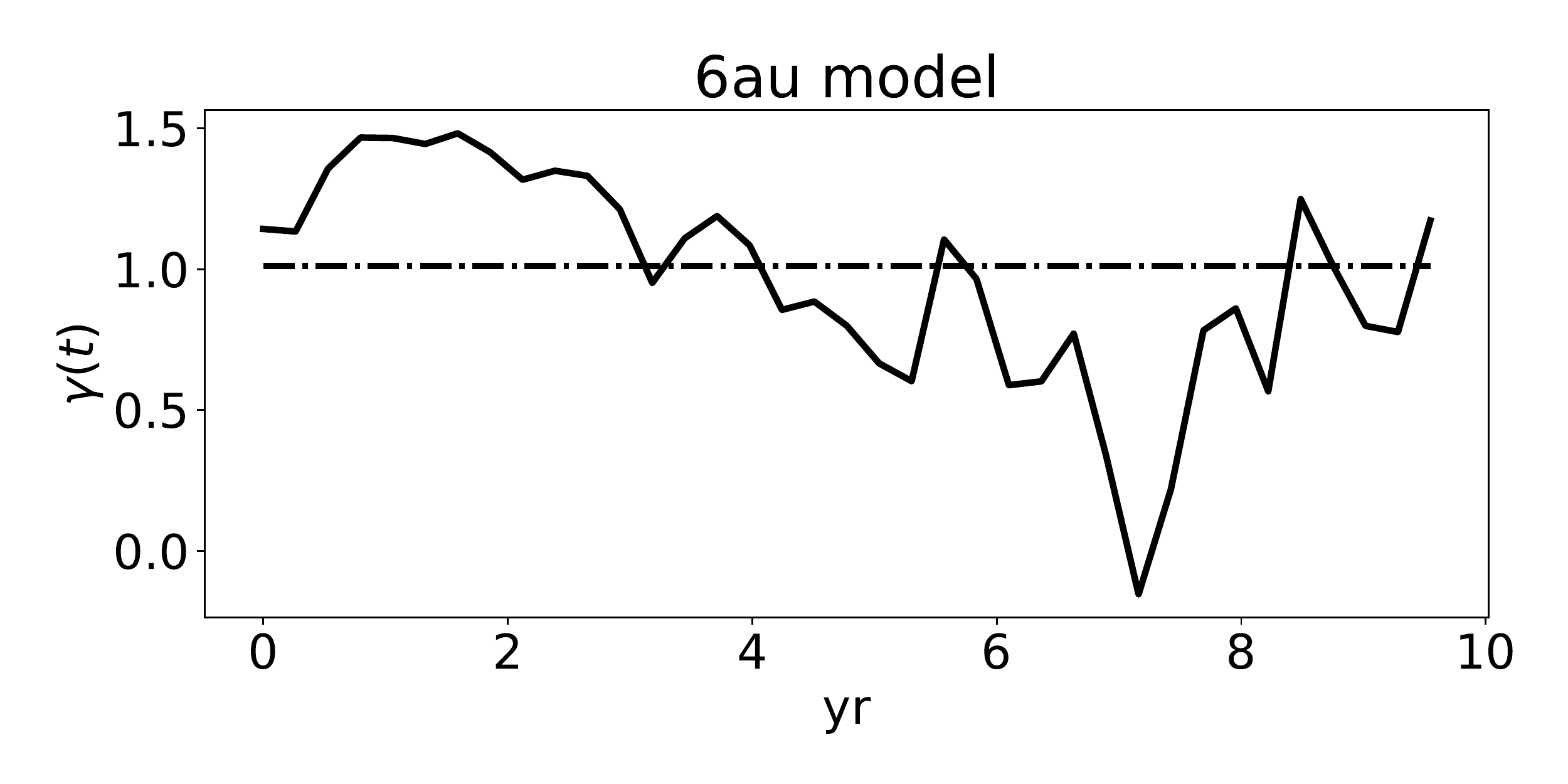}\\
    \includegraphics[width=1.0\columnwidth]{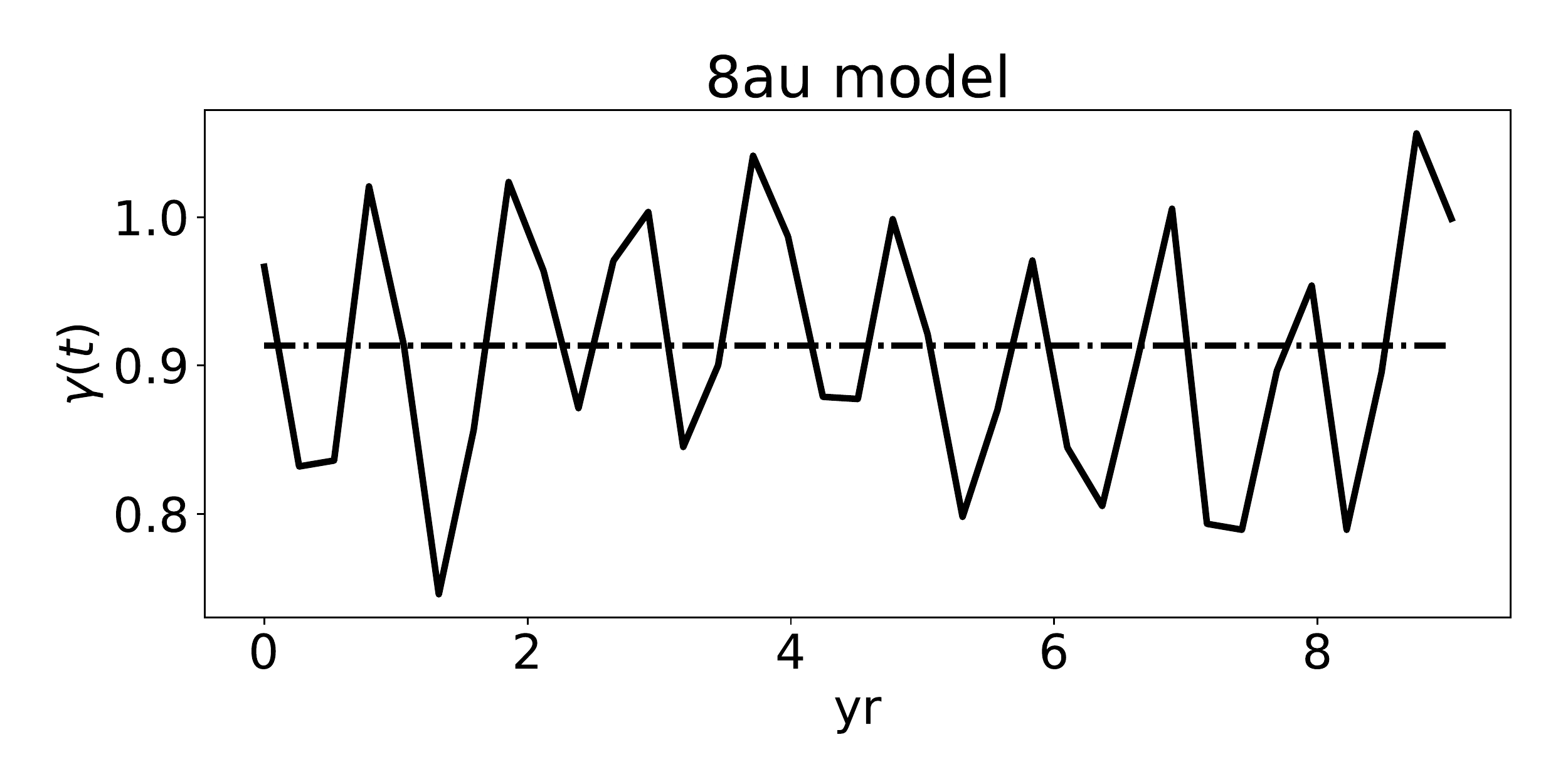}\\
    \includegraphics[width=1.0\columnwidth]{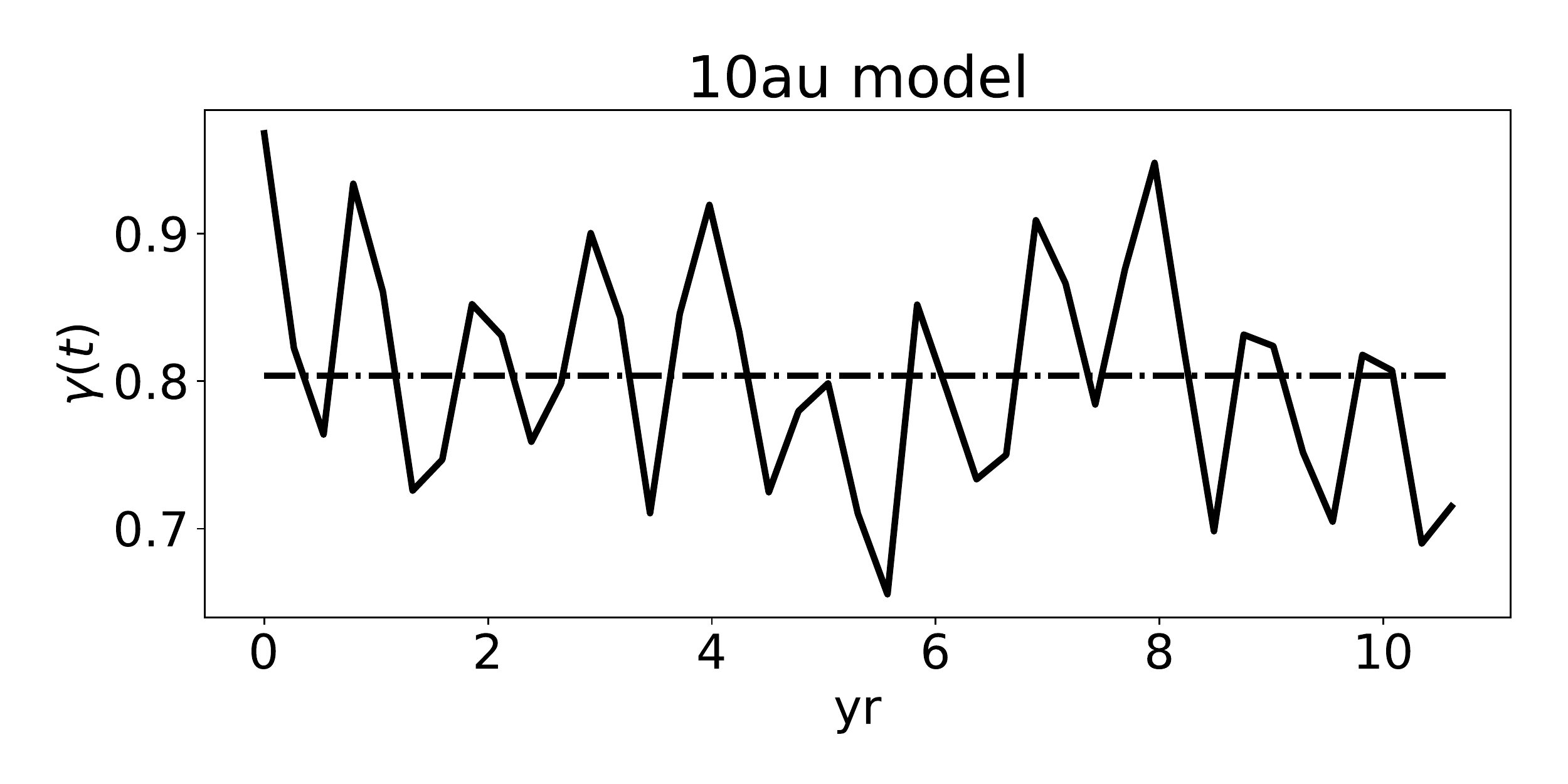}
    \caption{The solid lines show the sampled time varying $\gamma(t)$ of our five binary models. The dashed lines are the averaged $\gamma$.}
    \label{fig:gamma}
\end{figure}

Fig. \ref{fig:gamma} shows the time varying $\gamma(t)$ and averaged $\gamma$ of each models. There is a $1$yr period in $\gamma(t)$ corresponding to  the AGB pulsation. We also notice that there are dips in the $4$au and $6$au models but not in other models. We infer these to be  waves in the accretion disc and fall back. Both $4$au and $6$au simulations have large accretion discs (see Fig. 4 in \citet{chen2017}). Since the sampling shell is at $r_\text{flux}$, the waves in the accretion disc and also in the circumbinary disc can propagate to the sampling shell. However, we do not have a self-consistent model of these waves  (see discussions in Sec. 2.3 in \citet{chen2017}) for the accretion disc and   view this problem as a  future research direction.

\bsp	
\label{lastpage}
\end{document}